# Strain Engineering 2D MoS$_2$ with Thin Film Stress Capping Layers

Tara Peña,*,† Shoieb A. Chowdhury,‡ Ahmad Azizimanesh,† Arfan Sewaket,†

Hesam Askari,‡ and Stephen M. Wu*,†,¶

†Department of Electrical & Computer Engineering, University of Rochester, Rochester, NY, USA.

‡Department of Mechanical Engineering, University of Rochester, Rochester, NY, USA.

¶Department of Physics & Astronomy, University of Rochester, Rochester, NY, USA.

E-mail: tpena@ur.rochester.edu and stephen.wu@rochester.edu

## Abstract

We demonstrate a method to induce tensile and compressive strain into two-dimensional transition metal dichalcogenide (TMDC) MoS$_2$ via the deposition of stressed thin films to encapsulate exfoliated flakes. With this technique we can directly engineer MoS$_2$ strain magnitude by changing deposited thin film stress, therefore allowing variable strain to be applied on a flake-to-flake level. These thin film stressors are analogous to SiN$_x$ based stressors implemented in industrial CMOS processes to enhance Si mobility, suggesting that our concept is highly scalable and may be applied for large-scale integration of strain engineered TMDC devices. We choose optically transparent stressors to allow us to probe MoS$_2$ strain through Raman spectroscopy. Combining thickness dependent analyses of Raman peak shifts in MoS$_2$ with atomistic simulations, we can explore layer-by-layer strain transfer. MoS$_2$ on conventional substrates (SiO$_2$, MgO) show strain transfer into the top two layers of multilayer flakes with limited strain transfer to monolayers due to substrate adhesion. To mitigate this limitation, we also explore stressors on van der Waals heterostructures constructed of monolayer (1L) $MoS_2$ on hexagonal boron nitride (h-BN). This concept frees the 1L-$MoS_2$ allowing for a 0.85% strain to be applied to the monolayer with a corresponding strain induced bandgap change of 75 meV. By using thin films with higher stress, strain may be engineered to be even higher. Various stressors and deposition methods are considered, showing a stressor material independent transfer of strain that only depends on stressor film force with negligible defects induced into $MoS_2$ when thermal evaporation is used.

Transition metal dichalcogenides (TMDCs) have a multifaceted library of properties that can be altered with external perturbations, such as electric field effect, doping, etc.[1] Since these materials also exhibit high elastic limits, there exists an opportunity to use strain as another degree of freedom in engineering new nanoelectronic devices.[2] Strain engineered TMDCs may lead to a new generation of devices that utilize a higher degree of control over structural, electronic, optical, magnetic, superconducting, and topological materials' properties, which may now be controlled through strain.[3–7] Strain engineering techniques have already been well explored and implemented in semiconductor manufacturing, where silicon-based transistors are strain engineered for higher mobility through lattice mismatched epitaxial growth or deposition of thin film stressors such as $SiN_x$.[8,9] These concepts are so ubiquitous that almost all electronics today involve some degree of strain engineering, since the nanofabrication process itself will always create some amount of process induced strain that may be freely used to enhance device performance. These strain engineering techniques, therefore, are well-characterized for 3D bonded materials. However, there exists a gap to understand the implementation of these techniques onto 2D-bonded systems where the main feature is weak out-of-plane mechanical coupling. Additional considerations may need to be considered for strain engineering 2D systems due to the strongly anisotropic nature of the material, which may also have unique nanoscale materials properties such as interlayer slippage, thickness dependent mechanical properties, and variable substrate adhesion.

In this work, we explore strain engineering 2D systems through the deposition of stressed thin films onto exfoliated $MoS_2$ multilayer and monolayer flakes. Using this concept, which has been highly popular in 3D-bonded strain engineering, we are able to directly show through Raman spectroscopic mapping that we can directly engineer the strain state within our 2D TMDC. This technique is unique because we are able to engineer either **tensile** or **compressive** strain using the same process, and the magnitude of this strain may be directly controlled by the magnitude of the film force (film stress ($\sigma_f$) · film thickness ($t_f$)). Similarly, since stressed thin films may be deposited on a device-to-device basis, we are able to strain engineer each individual flake separately using this highly scalable technique that has already been heavily applied for the large-scale integration of Si based electronics. By enabling variable control of strain in individual 2D TMDCs directly on-chip, we may enable an entirely new domain of densely integrated 2D devices with strain engineered materials properties that has hitherto not been explored due to the few methods available for strain engineering 2D systems on-chip. Within this work, we will also specify the unique challenges to strain engineering 2D materials, such as out-of-plane strain transfer lengthscale, interlayer shear strength, substrate adhesion, engineering van der Waals (vdW) heterostructures, and minimization of defects induced by thin film deposition processes.

There have been several approaches in the past to strain 2D materials, specifically using lattice mismatched growth processes, fabricating suspended membrane structures, bending flexible substrates, and using diamond anvil cells.[10–13] While these approaches allow for control over strain, they are not ideal for highly integrated on-chip applications since they require complicated growth processes or external macroscopic mechanical forces. Moreover, the key difference is that many of these concepts apply strain **globally**, whereas in engineering TMDC-based devices the strain needs to be controlled **locally** at the individual flake level and applied at a large scale to many devices on a single chip. Imagining an integrated electronic or photonic circuit with heterogeneous 2D-TMDC based materials, the strain may not only need to be applied locally but the magnitude and nature (compressive or tensile) might need to be individually controlled for each separate material on-chip depending on application. While there has been progress in local strain application by the community,[14-16] we present a slightly different mechanism where strain can be transferred by taking advantage of evaporated stressed polycrystalline thin films. Our technique has the advantage of pre-existing wide scale adoption in past and current strained Si processes in CMOS manufacturing.

We confirm such a tunable strain process in exfoliated $MoS_2$ flakes, requiring only one evaporation process to encapsulate the 2D material with a stressed thin film. These evaporated stressors are polycrystalline thin films, which exhibit process induced stress similar to chemical vapor deposition (CVD) grown $SiN_x$ thin films. Almost all thin films exhibit process induced stress that results from its microstructural evolution during thin film growth.[17] Generally, tensile stress develops in low adatom mobility materials with faster growth rates, while compressive stress develops in high adatom mobility materials with slower growth rates.[18,19] Our work seeks to characterize tunable control of strain induced into 2D TMDCs from simple evaporation of dielectrics, as evaporation is inevitable in order to integrate 2D TMDCs on-chip. Evaporation of magnesium fluoride ($MgF_2$) is well-known to provide a tensilely stressed film, while other materials such as magnesium oxide (MgO) and silicon dioxide ($SiO_2$) provide compressively stressed films.[20–22] Fig. 1a summarizes the samples with a visual representation, where evaporated optically transparent thin film stressors are deposited onto exfoliated $MoS_2$. For the case of depositing a tensile thin film stressor, the film attempts to relax into a zero stress state by contracting. Therefore, after depositing a tensile thin film stressor onto $MoS_2$, the stressor will contract and therefore lead to compressive strain transferred into the $MoS_2$ layers. Similarly, with a compressive thin film stressor, the stressor will lead to tensile strain transferred into the $MoS_2$ layers. We specifically choose $MoS_2$ as the base material for our first demonstration, since the Raman modes with respect to strain and doping are characterized thoroughly.[11,23] We emphasize that this technique is meant to apply to any 2D vdW bonded material.

## Results & Discussion

As the simplest demonstration, we first examine the strain transferred from stressors into $MoS_2$ exfoliated onto MgO single crystal substrates, we later demonstrate this with $MoS_2$ exfoliated onto $SiO_2$/Si substrates. Care is taken with substrate surface pre-preparation to promote adhesion between the MgO substrate and the exfoliated MoS₂ flakes (see Methods). Post-exfoliation, optically transparent thin film stressors are deposited on our MoS₂/MgO samples through e-beam evaporation (Fig. 1a). We choose to evaporate $MgF_2$, MgO, and $SiO_2$ as the stressor layers since these coatings develop reliable and reproducible tensile or compressive stress. In our first set of experiments, we use a multilayer stressor of $Al_2O_3$ (10 nm)/X/$Al_2O_3$ (10 nm), where X = $MgF_2$, MgO, $SiO_2$. We refer to this trilayer geometry as a "multilayer X stressor" for brevity throughout the rest of this work. The bottom $Al_2O_3$ promotes adhesion at the stressor/substrate interface to prevent delamination, verified through microscratching tests.[24] More importantly, we employ the $Al_2O_3$ layer initially to keep the same interface for all the samples and focus on effects of varying thin film force only. It is well-known that humidity exposure may alter the stress within the film, and we find that the deposition of a thin non-porous amorphous top capping layer will protect the stressor material from this particular type of relaxation by forming a dense humidity proof cap.[25] Thin film force is defined as thin film stress times thin film thickness ($\sigma_f \cdot t_f$), which is the measurement that quantifies the load being applied onto MoS₂ from the stressors. The film force within these evaporated stressor films may therefore be simply controlled by adjusting the film thickness. We have observed highly reproducible thin film stress and force from $MgF_2$, MgO, and $SiO_2$ when the films are deposited under the same evaporation conditions (see Methods for evaporation details, Fig. S5). We determine the stress distributions from the deposited thin film stressors to be biaxial through our wafer curvature measurements, matching the expected results from all other works.[26-28] By applying a biaxial stressor uniformly onto our flakes, it is likely that biaxial strain is transferred, as this has been the observed result in other uniform coverage works with $SiN_x$.[29-32] Spatial Raman maps were collected for exfoliated MoS₂ samples with stressed encapsulations varying in thin film force (-30 N/m to +30 N/m), and we specifically study MoS₂ flakes with thicknesses ranging from 1L-7L. We focus on the $E^1_{2g}$ peak to characterize strain in the MoS₂ samples, since this peak has been experimentally and theoretically proven to be the most sensitive to in-plane biaxial strain.[11]

Fig. 1b presents Raman signatures from bilayer (2L) samples with varying thin film forces extracted from the spatial Raman mappings. A clear shift is seen in the $E^1_{2g}$ peak position, which is expected when strain is transferred into the $MoS_2$ samples. **Tensile** multilayer $MgF_2$ stressor films (red) create in-plane **compressive** strain throughout the $MoS_2$ samples, leading to a positive $E^1_{2g}$ peak shift of ~2 cm$^{-1}$. **Compressive** multilayer MgO stressor films (blue) create in-plane **tensile** strain throughout the $MoS_2$ samples, leading to a negative $E^1_{2g}$ peak shift of ~2 cm$^{-1}$. To demonstrate this point further, Raman mappings are presented before and after stressor encapsulation on the same 2L samples. This map shows a spatial distribution of $E^1_{2g}$ peak position on each flake before and after either a tensile (Fig. 1c-f) or compressive (Fig. 1g-j) thin film is deposited. We observe a spatially uniform ~2 cm$^{-1}$ increase in the $E^1_{2g}$ peak position from before encapsulation (Fig. 1e) to after tensile multilayer $MgF_2$ stressor encapsulation (Fig. 1f). Similarly, with our compressive multilayer MgO stressor thin films, we observe a spatially uniform decrease in the $E^1_{2g}$ peak position of ~2 cm$^{-1}$ from before encapsulation (Fig. 1i) to after encapsulation (Fig. 1j).

These results represent a direct demonstration of our thin film stressor induced strain concept, and shows that we can induce either tensile or compressive strain using thin film stressors on the flake to flake level. As the same $Al_2O_3$ layer initially contacts the $MoS_2$, it is unlikely that both upward and downward shifts in the $E^1_{2g}$ peak position can be explained by damage/defects. The peak shifts are only correlated with positive/negative (tensile/compressive) film force of the X layer in the stressor (X = $MgF_2$ or MgO), strongly suggesting that these results represent in-plane strain in the $MoS_2$ layer. More detailed analysis of defects induced by the deposition process, as well as a combined analysis of $E^1_{2g}$ and $A_{1g}$ peak shifts will be analyzed later when we examine the strain effect on monolayer $MoS_2$.

To understand the effect of $MoS_2$ layer thickness on strain transfer, Raman mappings were conducted for encapsulated and non-encapsulated flakes from 1L to 7L. Fig. 2a presents the $E^1_{2g}$ peak shifts of $MoS_2$ flakes with a tensile multilayer $MgF_2$ encapsulation (red), no encapsulation (grey), and a compressive multilayer MgO encapsulation (blue). The $E^1_{2g}$ peak position of the stressed samples begin to clearly diverge after 4-6 layers and below, hinting that the strain transferred into our samples may not be uniformly distributed throughout the flake in the out-of-plane direction. We have also replicated these results with $MoS_2$ flakes that are exfoliated onto $SiO_2$/Si substrates (Fig. S13), suggesting that all of our results are equally applicable to $MoS_2$/$SiO_2$/Si samples. It is likely that the strain is localized to the top few layers of $MoS_2$ due to the weak out-of-plane mechanical coupling in 2D materials. We also present the difference in peak position relative to the control for each thickness in the stressed samples for clarity (Fig. 2b). It is important to note that in encapsulated monolayer (1L) samples, there is only a small amount of strain transferred. This small strain transfer is an indication that the bottom layer is fixed to the substrate. This fixed boundary condition relates to the adhesion at the $MoS_2$/substrate interface, where delamination of the 2D material from the substrate will occur when stressors are deposited without proper substrate adhesion. To further prove the robustness of our stressor induced strain transfer concept, we present the peak shifts for each thickness flake while varying thin film force (Fig. 2c). The results indicate a direct linear trend between film force and peak shifts, re-confirming that the strain transferred into the $MoS_2$ flakes originates from application of thin film force and nothing else. This is not only a direct demonstration that our strain effects come from the stressor layers, but also that this strain transfer effect is completely tunable with only one parameter, film force. By knowing only the film force deposited onto $MoS_2$ we are now able to directly control the strain state of our 2D material. The slope of this linear coupling of peak shift and film force are extracted for each thickness and presented in Fig. 2d, again displaying an exponential dependence with respect to MoS$_2$ layer thickness. The exponential dependence hints that there is a heterogeneity of strain throughout each of the layers, exactly as the results from Fig. 2a,b suggest. In the presence of heterostrain, we must acknowledge the overall Raman signal being the result of superimposed Raman signatures from each layer within a given sample. In the case of a 2L sample, the top layer will be strained entirely by the stressor while the bottom layer is to a degree fixed to the substrate. Therefore, the measured Raman signal for the 2L sample is the coexistence of Raman signatures from the top strained layer and bottom fixed layer (with little to no strain). We next combine these results with computational simulations to confirm the actual strain distribution in the c-axis.

## Molecular Statics Simulations & Comparison to Experimental Data

To fully understand the layer-by-layer nature of the strain transferred into our 2D material from the stressors, molecular statics (MS) calculations (see Methods) were conducted for MoS$_2$ samples (2L-7L). A many-body reactive empirical bond-order (REBO) potential was used to model the covalent interactions and a two-body Lennard-Jones potential was used for interlayer vdW bonds.[33,34] This specific potential has been widely reported to accurately predict structural and mechanical properties, in addition to simulating structural phase transformations and complex mechanical loading (i.e. nanoindentation).[35,36] Given that all of our samples experimentally are fully encapsulated, we choose to mimic this in simulations with an in-plane biaxial strain distribution to the top MoS$_2$ layer (to replicate pulling in all directions). We choose MS to extract the strain distributed throughout each layer since there are two imperative interfaces to control, strain being applied at the stressor/MoS$_2$ interface and whether the bottom layer of MoS$_2$ is fixed or free to the substrate. While in an ideal system, the MoS$_2$/substrate interface would be perfectly fixed, we do observe a small amount of strain transferred into our 1L samples (Fig. 2a). To emulate a "mostly" fixed boundary condition in our simulations, we compute the strain within each layer (for samples of 2L-7L) for both fixed and free boundary conditions. To obtain a final result for the strain in each layer, we take a weighted average of the two results for each layer within a given sample. The weighting factor that best matches experimental results was the 75% fixed and 25% free. Fig. 3a shows a visual representation of the simulated results, while Fig. 3b shows the exact computationally simulated strain within each layer for 2L-7L samples with a 75% fixed boundary condition. The 2L sample has 0.85% strain concentrated within the top layer, while only a small amount of strain transferred of 0.07% to the bottom layer for this sample since the bottom layer is predicted to be 75% fixed. For samples 3L-7L, the top layer has a strain of 0.85% while the next layer decreases to 0.12%, then the bottom layers have negligible amounts of strain.

To directly compare what we see experimentally to the computational results, we first quantify translation factors to convert Raman peak shifts (cm$^{-1}$) to strain (%) (Fig. S6). The translation values are used are from previous biaxial strain work.[11,12] The overall Raman signature measured experimentally is a superposition of optical responses from each layer within the sample, similar to what is observed in Raman on TMDC heterostructures.[37,38] Experimentally these peaks are superimposed into one single peak, which is attributed to acknowledging the Gaussian linewidth from the instrumental response being on the order of the peak shifts. In attempts to deconvolve these peaks, we calculate the $E^1_{2g}$ Lorentzian responses from each layer (for 2L-7L samples), where the peak position of each layer is set to match the strain based off computational results. We superimpose the responses from each layer for each sample thickness, then extract the peak position of the resulting response (Fig. S7). Other parameters used in this calculation such as layer response intensities and full-width-half-maximums are extracted from experimentally measured data. Upon comparing the calculated peak shifts to that of our experimental results, we confirm the exponential decay trend of $\Delta E^1_{2g}$ matches what we found experimentally (Fig. 3c). Note in Fig. 3c, we directly compare both tensilely and compressively strained samples by taking the absolute values of $\Delta E^1_{2g}$. Strain transferred into the first two layers matches the strain penetration regime observed in other works, where strain is induced into $MoS_2$ from silver nanoparticles.[39]

From our analysis, it is clear that there is a strain transfer lengthscale in the out-of-plane direction for $MoS_2$. We attribute this to incomplete transfer of shear traction between $MoS_2$ layers due to weak interlayer bonding. Shear-lag models have successfully quantified strain transfer properties between graphene and flexible substrates, typically employing uniaxial tension onto the 2D material.[40] Models typically include terms to account for incomplete transfer of traction between 2D materials and various substrate materials under applied force, governed by an interfacial stiffness constant. This interfacial stiffness constant describes the linear relation between traction and displacement, and variations in this constant between various 2D materials depending on the strength of interlayer interaction may explain the nature of our observed out-of-plane strain transfer lengthscale. Shear-lag models have also been used to quantify critical values of interfacial shear strength where layers may begin to slip. Work from Kumar *et. al*. has explored 2L vdW heterostructures to quantify critical strains, where layer slipping may occur and generate strain solitons.[41] The critical strain within these systems is strongly correlated with the interlayer shear strength. Kumar *et. al*. finds 2L graphene and 2L $WSe_2$ systems with the smallest and largest critical strain (interlayer shear strength) respectively. Levita *et. al.* observes a similar result by calculating the work of separation between 2L systems, where 2L graphene and 2L $MoTe_2$ have the smallest and largest work of separation respectively.[42] For our experiments, we choose to stay below any predicted interlayer slippage that may occur in $MoS_2$, which has been suggested in literature to be around 1.8% strain. This is supported by our MS simulations, where we observe no strain induced interlayer slippage at the magnitudes we consider. Our results here suggest that engineering the strain penetration depth in the c-axis with our stressor induced strain transfer technique strongly depends on the interlayer properties within a given 2D system.

The interlayer properties of the 2D system may be predetermined by the material for single composition 2D systems, but may also be engineered using vdW heteroepitaxy.

## Strain Engineering van der Waals Heterostructures

Many 2D materials have exotic properties when thinned down to the 1L regime that may be manipulated with strain. While we have observed robust tunable strain for few-layer $MoS_2$ on a 3D bonded substrate, we have not been able to strain 1L-$MoS_2$ due to its fixed nature at the $MoS_2$/substrate interface. We choose to overcome this limitation by exploring the replacement of a normal 3D bonded substrate ($SiO_2$ or MgO) with a 2D weakly bonded out-of-plane material such as hexagonal boron nitride (h-BN).

To explore stressor effects on vdW heterostructures, we use a dry transfer technique (see Methods) to construct a 1L-$MoS_2$/h-BN heterostructure on a conventional $SiO_2$/Si wafer. We perform Raman mappings before and after tensilely stressed thin film stressor encapsulations on 1L-$MoS_2$/h-BN/$SiO_2$/Si samples to test the possibility of straining 1L-$MoS_2$. To ensure our technique is completely robust to multiple types of stressor compositions and also to multiple different stressor deposition techniques, we explore the use of two different stressor layers. First, is the conventional e-beam evaporated multilayer $MgF_2$ stressor that we have considered before with top and bottom $Al_2O_3$ capping layers (Fig. 4a-c). Second, is a thermally evaporated $MgF_2$ alone stressor (Fig. 4d-f). We first test a sample encapsulated with the typical e-beam evaporated $MgF_2$ multilayer described in Fig. 1-2. The average $E^1_{2g}$ peak position shift before encapsulation (Fig. 4b) to after encapsulation (Fig. 4c) was an increase of 3.8 cm$^{-1}$ (0.73% compressive strain). The maximum $E^1_{2g}$ peak shift we observe in this sample is as high as 4.4 cm$^{-1}$ (0.85% compressive strain), which is the strain regime we anticipated from previous samples and used in our MS computational results. Then, Raman mappings before and after encapsulation were also tested again for the thermally evaporated $MgF_2$ alone film (Fig. 4e,f), where an average $E^1_{2g}$ peak position shift of 1.7 cm$^{-1}$ (0.34% compressive strain) and maximum $E^1_{2g}$ shift of 1.9 cm$^{-1}$ (0.37% compressive strain). The smaller strain transferred into the 1L-MoS$_2$ can be attributed from the differences of the thin film force between the two films (16 N/m and 25 N/m). This confirms that we are able to directly strain 1L-MoS$_2$ when using a 1L-MoS$_2$/h-BN heterostructure, and that the result does not depend on any other parameter besides film force. This effect is stressor composition independent, as we see that both stressor films equally strain 1L-MoS$_2$ with a magnitude proportional to its film force. One advantage of using 1L-MoS$_2$ is that Raman intensity ratio of the two dominant peaks ($E^1_{2g}$ and $A_{1g}$) may also be used as an independent confirmation of biaxial strain, and we also find a linear relationship of this ratio to film force (Fig. S8).[43] Spatially averaged Raman spectra from these samples are presented in Fig. 4g, in addition to photolumienscence (PL) signatures in Fig. 4h. The A exciton peak with biaxial strain should shift ~100 meV/%, which we extract a shift of 75 meV and 35 meV for the higher and lower thin film force samples respectively (as expected). In the inset plot of Fig. 4g, and Fig. 4h, we also see a linear relationship between both Raman peak shift and exciton peak shift in the 1L-MoS$_2$ samples, which provides further independent proof that our strain transfer technique works and is only dependent on film force.

## Comparison of Different Stressor Deposition Techniques

We next conduct a thorough characterization of various types of stressor depositions on 1LMoS$_2$/h-BN/SiO$_2$/Si samples, to characterize if defects are induced into the MoS$_2$ through these deposition processes. It is well-known that various methods of thin film deposition may create defects and disorder when deposited on 2D materials, with thermal evaporation typically considered the best method due to the well thermalized nature of the deposited thin film material.[44] To test these effects out we choose to examine four different stressors: the previously explored e-beam evaporated multilayer Al$_2$O$_3$/MgF$_2$/Al$_2$O$_3$ (25 N/m), thermally evaporated

CrO$_x$/MgF$_2$ (20 N/m), thermally evaporated TiO$_2$/MgF$_2$ (12 N/m), and thermally evaporated MgF$_2$ alone (16 N/m). MgF$_2$ layers were all kept at 100 nm thicknesses. The Al$_2$O$_3$ and TiO$_2$ capping layers presented were kept at 10 nm in thickness while the CrO$_x$ was kept at 5 nm, they were implemented to protect the stressors from humidity effects. We plot the Raman spectra of the three encapsulated samples and a control (no encapsulation) sample (Fig. 5a). We next extract the full-width-half-maximum (FWHM) of the A$_{1g}$ peak, $\Gamma$(A$_{1g}$), before and after encapsulation for the three 1L-MoS$_2$/h-BN/SiO$_2$/Si samples (Fig. 5b, see Fig. S9 for $\Gamma$(E$^1_{2g}$ )). Both thermally evaporated CrO$_x$/MgF$_2$ and MgF$_2$ alone encapsulated sample has negligible changes in the $\Gamma$(A$_{1g}$), however $\Delta\Gamma$(A$_{1g}$) is 0.84 and 3.61 cm$^{-1}$ for thermally evaporated TiO$_2$/MgF$_2$ and e-beam evaporated MgF$_2$ multilayer encapsulated samples respectively. FWHM is highly sensitive to the presence of disorder, increasing upon further introduction of disorder in the system.[45,46] Examination of individual spectroscopic traces in our Raman map indicate that our FWHM changes are due to homogeneous broadening by materials changes and not an effect of spatially averaged inhomogeneous broadening due to the A$_{1g}$ peak variation. Within our experimental determination both thermally evaporated CrO$_x$/MgF$_2$ and MgF$_2$ alone stressors do not increase the FWHM, indicating that we can achieve the effects of stressor induced strain transfer **without** any defects being introduced into the 2D system. We reconfirm that this thermally evaporated MgF$_2$ alone stressor behaves the same when used on multilayer MoS$_2$ samples, in the same experiment that we conducted in Fig. 2 with the e-beam evaporated stressors (Fig. S10). We observe the exact same exponential decay in the E$^1_{2g}$ peak shift with respect to MoS$_2$ thickness that we observed before in Fig. 2. Defects may still be introduced in our TiO$_2$/MgF$_2$ films deposited through thermal evaporation, as higher currents are needed to thermally deposit TiO$_2$ which leads to increased sample heating during deposition.

To further evaluate the effect of disorder and defects, as well as the origin of the Raman peak shifts in our monolayer MoS$_2$ samples, we construct a $\Delta$A$_{1g}$ versus $\Delta$E$^1_{2g}$ plot for the given 1L-MoS$_2$/h-BN/SiO$_2$/Si encapsulated samples (Fig. 5c). These maps are calculated from the peak position differences from the before and after encapsulation Raman maps of our vdW heterostructures (Fig. 4b,c and Fig. 4e,f). Phonon frequency shifts for both E$^1_{2g}$ and A$_{1g}$ modes are not strictly functions of only strain, but also of charge carrier concentration. Doping changes the electron-phonon coupling for the Raman in MoS$_2$, which has been thoroughly observed in other works on electron doping trends with Γ and position of the A$_{1g}$ mode (electron-phonon coupling is greater for A$_{1g}$ than E$^1_{2g}$).[47] Therefore, it is necessary to separate the effect of strain and the effect of carrier concentration change that may also arise from either defect induced doping or band bending as a result of dielectric encapsulation.[48] Using a standard linear transformation, from the Grüneisen parameter of each phonon mode, and the corresponding phonon shift rate with carrier concentration, we can map a new strain-carrier concentration vector space onto our ΔA$_{1g}$ versus ΔE$^1_{2g}$ plot (Fig. 5c).[23] First, we examine the thermally evaporated MgF$_2$ only encapsulation layer. The mean ΔA$_{1g}$ position is as little as -0.01 cm$^{-1}$, for these heterostructures, corresponding to a Δn ~ 0.01×10$^{13}$ cm$^{-2}$. This small change in carrier concentration may not arise from defects, but more likely is due to changes in the dielectric environment, which may change effects such as interfacial band bending in 1L-MoS$_2$. This hypothesis is supplemented by our previous result where only small changes occur to the FWHM of the A$_{1g}$ peak in these samples. We can also extract that we are applying ~0.33% compressive strain using the thermally evaporated MgF$_2$ only stressor, which scales with the reduced film force of this film (16 N/m) when compared with e-beam evaporated multilayer stressors from previous sections (25 N/m). Next, we examine the thermally evaporated CrO$_x$/MgF$_2$ film, where the mean ΔA$_{1g}$ is -0.02 cm$^{-1}$. This is similar behavior to the thermally evaporated MgF$_2$ alone encapsulation, there is still likely due to changes in the dielectric environment. As the thin film force of the CrO$_x$/MgF$_2$ is 20 N/m, the sample exhibited a mean E$^1_{2g}$ shift of 2.46 cm$^{-1}$ (0.47% compressive strain) and maximum E$^1_{2g}$ shift of 2.77 cm$^{-1}$ (0.53% compressive strain). Finally, we examine the thermally deposited

TiO$_2$/MgF$_2$ stressor. The mean $\Delta A_{1g}$ position for these samples reaches -0.47 cm$^{-1}$, corresponding to a $\Delta n \sim 0.2 \times 10^{13}$ cm$^{-2}$. Since the dielectric environment between the two previous samples are similar, it is likely there are defects or disorder introduced into the system that causes low level doping that changes carrier concentration. This is supported by the minor increase in the FWHM from the previous section. The 1L-MoS$_2$ is strained to a similar degree as the previous sample, due to the fact that the film forces are of similar magnitude. Finally, the e-beam evaporated MgF$_2$ encapsulated sample has a mean $\Delta A_{1g}$ of 1.28 cm$^{-1}$, the increases in the $A_{1g}$ peak position along with the increase in $\Gamma(A_{1g})$ suggests there may be more disorder introduced into the MoS$_2$. At higher defect concentrations a unique disorder induced Raman mode appears, and defect density may be quantified with the intensity of this LA(M) peak at $\sim$227 cm$^{-1}$. Upon further investigation into the Raman spectra of the e-beam evaporated MgF$_2$ multilayer encapsulated sample, there is an observable LA(M) peak present, which is absent in the other two samples (Fig. S11). The interdefect length ($L_D$) may be quantified by observing the intensity of the LA(M) with respect to the intensities of and $A_{1g}$ individually, therefore we calculate a $L_D$ of 3.9 ± 1 nm for this sample.[45] Using this value of the $L_D$, we can calculate an expected change in carrier concentration, if each defect serves as a p-type dopant, of $\Delta n \sim$ -0.65$\times 10^{13}$ cm$^{-2}$ where $\Delta n$ is the change in electron concentration. This result matches well to the expected $\Delta n$ when extracting from the $\Delta A_{1g}$ versus $\Delta E^1_{2g}$ plot. The presented e-beam evaporated MgF$_2$ multilayers have a higher thin film force than the thermally evaporated films presented, which is seen in transferred strain into the MoS$_2$ ($\Delta E^1_{2g}$) in Fig. 5c (also in Fig. 4g,h). While the e-beam evaporated MgF$_2$ multilayers hold more stress and transfers strain up to -0.85%, this is at the expense of introducing more disorder. Conversely, thermally evaporated thin film stressors are confirmed to introduce no observable amount of doping/disorder and strain up to -0.53% when implementing the thermally evaporated CrO$_x$/MgF$_2$ stressor. CrO$_x$/MgF$_2$ is the most thermal, time, and humidity stable thin film stressor we present. While we only demonstrated a 20 N/m application, one may engineer this thin film stressor to have a greater thin film force by adjusting thickness/growth rate parameters. Spatial uniformity of strain applied to the flake for these different thin films are also discussed in the supplementary information (Fig. S12). Thin film deposition techniques are important to consider when strain engineering realistic systems for applications, although some of these effects may be mitigated by introducing a predeposited protective capping onto 1L-MoS$_2$ before stressors are applied. Protective capping layers have been highly successful in damage reduction in 2D TMDCs.[49,50]

## Conclusion

We have been able to show that with the deposition of thin film stressors, we are able to strain engineer 2D MoS$_2$. By controlling the film force, we are directly able to tune the amount of either tensile or compressive in-plane strain in our MoS$_2$ flakes. For the magnitudes of strain this study, the strain has been shown to penetrate two layers in the c-axis of multilayer flakes. The strain transfer lengthscale is likely unique to each 2D material depending on interlayer vdW coupling, with MoS$_2$ and graphene being close to the lower limit based on experimental and theoretical findings.[42] Our findings match with other experimental demonstrations where strain transfer is related to interlayer adhesion, where the degree of strain transfer decay depends on the strength of the vdW bond.[51] Other TMDCs such as MoSe$_2$ and MoTe$_2$ have been predicted to have improved interlayer adhesion correlating with the size of the chalcogen, therefore they have a strong possibility of exhibiting longer strain penetration depths. Using these ideas, we have also shown that utilizing vdW heteroepitaxy can overcome substrate adhesion concerns when straining monolayer TMDCs. Using dry transferred monolayer MoS$_2$ on h-BN we are able to directly apply large strains into the monolayer material with minimal defects using thermally evaporated stressor materials. A detailed study of damage induced by the stressor also has shown that thermally evaporated stressors induce negligible defects compared to e-beam deposition, this matches with what is understood about these deposition processes on 2D materials. Additionally, evaporation in general has been observed to induce less damage to 2D materials than processes such as sputtering, pulsed laser deposition, and etc.[44]

When adopting strain engineering techniques from standard CMOS technology onto 2D TMDCs, we have shown that weak interlayer bonding needs to be considered in understanding the strain transfer profile in the c-axis direction. After adjusting for these unique considerations between 2D and 3D bonded systems, the possibility opens for the large scale exploration of the unutilized strain degree of freedom in engineering 2D TMDC based devices. This is especially important for any process that may consider the heterogeneous large-scale integration of multiple TMDC based materials on a single chip. Our process can uniquely address such issues since the stressor layers with variable stress may be applied separately to each device or material in a densely-packed deeply-scaled integrated circuit. Since evaporation processes are accessible and common tools used in micro/nanofabrication, our highly tunable technique has large implications for engineering device structures with 2D materials. Nanopatterning these thin film stressors may allow for engineering specific strain patterns within device structures, this is a technique that has been utilized heavily in 3D bonded Si transistors.[52] Electric-field controllable dynamic strain from piezoelectrics has also been implemented onto 2D systems.[53,54] Our own work coupling thin film stress with dynamic strain from ferroelectric substrates has allowed for us to demonstrate a gate-controllable structural/electronic phase transition within $MoTe_2$.[55] Similarly, since interlayer bonding is uniquely weak in our materials, our technique represents one of the few ways to controllably apply heterostrain to multilayer 2D systems where each layer is independently strained with respect to each other. Heterostrain is an important but rarely explored method to engineer new quantum 2D materials from existing 2D systems, especially when related to 2D moirè superlattices in systems such as twisted bilayer graphene.[10] Strain engineering can be a powerful tool that may open the possibility for control over a wide-variety of strain-tunable material properties in 2D TMDCs. By adapting popular strain engineering techniques from 3D-bonded industrial processes, these ideas may be adopted on the device scale to create new and otherwise unobtainable functionality in novel 2D devices.

## Methods

The exfoliated control $MoS_2$ layers are characterized for thickness via Raman spectroscopy (Fig. S1), optical microscopy (Fig. S2), and photoluminescence spectroscopy (Fig. S3), with layer thicknesses verified with atomic force microscopy.[56,57] $MoS_2$ flakes are exfoliated from the bulk onto pre-polished single-crystal MgO substrates. The roughness of these substrates are confirmed to be 0.125 nm and below. We mitigate poor adhesion by annealing the substrates at 150°C in a humidity-controlled environment (<1 ppm $H_2O$ and $O_2$), limiting residual water and hydroxl groups on the substrates' surface prior to TMDC exfoliation.[58] Once exfoliation is completed, the samples are brought out of the glovebox and into an acetone, IPA ultrasound bath. The ultrasound bath allows for the poorly adhered flakes to be ripped from the substrate. The samples are then immediately placed into a vacuum chamber, where the thin film stressors are evaporated.

All deposition processes began at $5 \times 10^{-6}$ torr, with all growth rates kept between 1-2 Å/s. Cleaned coverslips are placed alongside of the samples during deposition. The radius of curvature of the glass slides are determined pre and post-deposition with a surface profilometer, which are used to determine thin film stress using the Stoney equation, the standard wafer curvature technique.[59] Using a WITec Alpha300R Confocal Raman Microscope, we perform Raman mapping with 250 nm step size resolution. The 532 nm laser was focused on the sample using a 100x objective (0.90 N.A.), the spotsize of the laser is estimated to be 0.7 $\mu m$. The power was monitored carefully to stay below 0.75 mW, to prevent sample damage from laser heating. Raman peak characterization is done by fitting to a Voigt spectral profile. Fitting to a Voigt allows one to extract the intrinsic phonon response, by compensating for the additional instrumental response (Fig. S4).

Atomistic modeling was carried out using the LAMMPS software package to study strain transfer on several $MoS_2$ structures with different number of layers.[60] Initial structures with almost square planar dimensions of 20 by 20 $nm^2$ were created, with each layer consisting of approximately 13,400 atoms and thickness was varied between 12.29 Å and 43 Å (corresponding to 2L-7L). All simulated $MoS_2$ samples follow a AA' stacking sequence with free surface boundary in all directions. Initially, structures were relaxed using a conjugate gradient energy minimization algorithm to ensure minimum energy configurations. Subsequently, the MS method was used (at T = 0 K) where a constant incremental biaxial strain in x and y direction was applied on the top layer of all the structures. Biaxial strain was incremented by Δε = 0.14% (ε = $\sqrt{\varepsilon_x^2 + \varepsilon_y^2}$) up to final biaxial strain magnitude of ε = 0.85%. Between each increment, the atoms at the top layer were kept stationary at the prescribed strain and energy minimization was performed. Finally, Ovito open visualization tool was used to visualize the atoms afterwards where atomic strains of all the atoms in a layer was used to compute the average strain of individual layers within a given structure (Fig. 3a).[61]

1L-$MoS_2$/h-BN/$SiO_2$/Si heterostructures are constructed with a typical dry transfer procedure. 1L-$MoS_2$ is confirmed via optical contrast and atomic force microscopy after exfoliation from the bulk onto an $O_2$ plasma cleaned 90 nm $SiO_2$/Si substrate. The 1L-$MoS_2$ is picked up using a dome shaped PC/PDMS stamp on a transfer stage, then it is dropped off on top of a large h-BN flake (~30-40 nm thickness) on a separate $SiO_2$/Si substrate by melting the PC at 200°C. The dome shaped PC/PDMS stamp is chosen to prevent bubble formations, then the sample is confirmed to be bubble-free with differential interference contrast microscopy. All exfoliations and transfer processes were conducted inside a humidity-controlled environment still. The PC is removed with chloroform, IPA baths to ensure a clean interface. The heterostructures from top to bottom follows 1L-$MoS_2$/h-BN/$SiO_2$/Si, ensuring the stressor will make contact with the $MoS_2$ upon deposition (Fig. 4a, d).

## Acknowledgment

We wish to acknowledge support from the National Science Foundation (OMA-1936250 and ECCS-1942815) and the National Science Foundation Graduate Research Fellowship Program (DGE-1939268). This work also made use of the Cornell Center for Materials Research Shared Facilities, which are supported through the NSF MRSEC programme (DMR1719875).

## Supporting Information Available

Presented after main text.

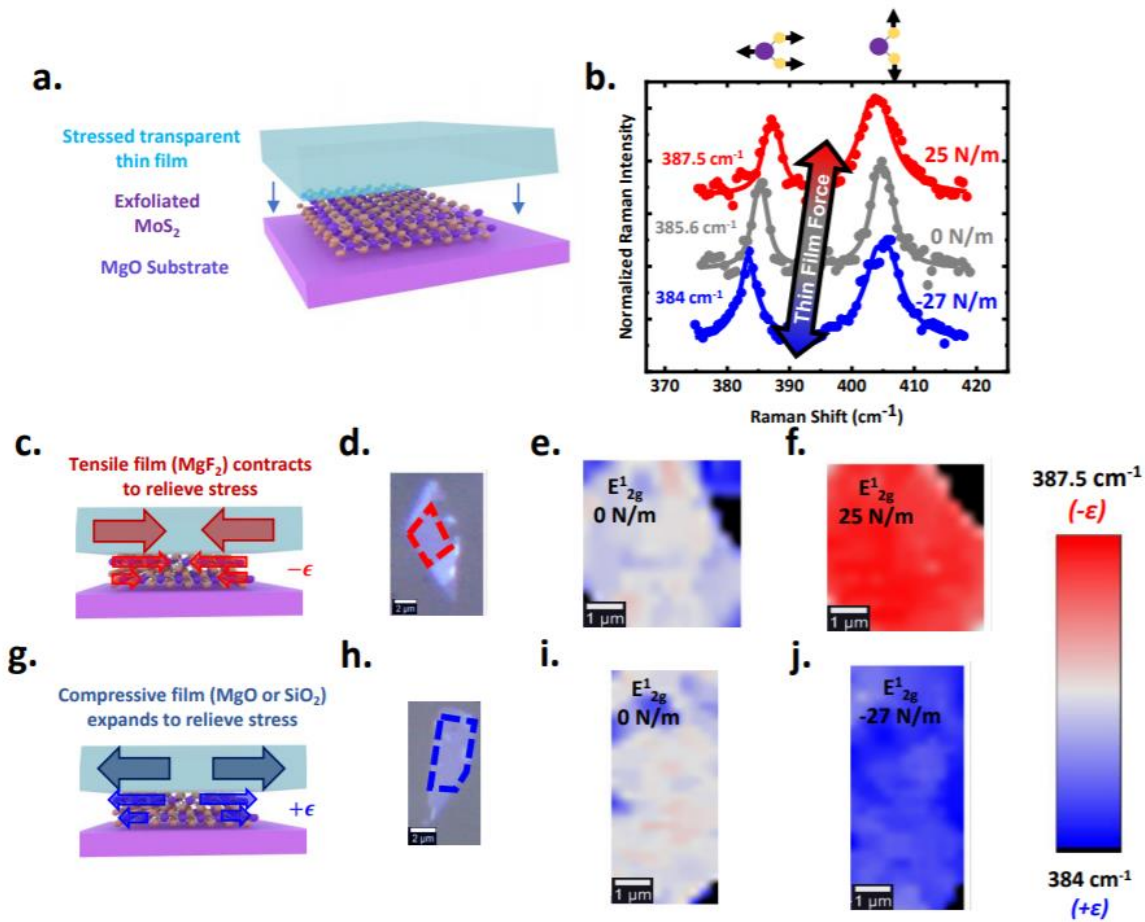

**Figure 1: (a)** Sample preparation. Evaporation of thin film stressors onto exfoliated MoS₂ samples. **(b)** Raman spectra of 2L MoS₂ samples while varying thin film force, demonstrating clear shifts in the $E^1_{2g}$ phonon mode. Red spectrum is that of a compressively strained MoS₂, grey is a control exfoliated MoS₂, and the blue is that of a tensilely strained MoS₂. **(c)** Visual representation of a tensile MgF₂ multilayer stressor contracting to release stress within itself, leading to compressively strained MoS₂. Strain transferred (smaller red arrows) is presented to vary layer-by-layer. **(d)** Optical micrograph of the 2L MoS₂ sample. **(e)** $E^1_{2g}$ peak position map of the flake from (d) with no encapsulation. **(f)** $E^1_{2g}$ peak position map of the flake from (d) after a tensile MgF₂ multilayer encapsulation. **(g)** Visual representation of a compressive MgO multilayer stressor expanding to release stress within itself, leading to tensilely strained MoS₂. Strain transferred (smaller blue arrows) is presented to vary layer-by-layer. **(h)** Optical micrograph of the 2L MoS₂ sample. **(i)** $E^1_{2g}$ peak position map of the flake from (h) with no encapsulation. **(j)** $E^1_{2g}$ peak position map of the flake from (h) after a compressive MgO multilayer encapsulation. All Raman maps presented here follow the color bar on the bottom right. Thin film stressors presented here are all e-beam evaporated multilayers.

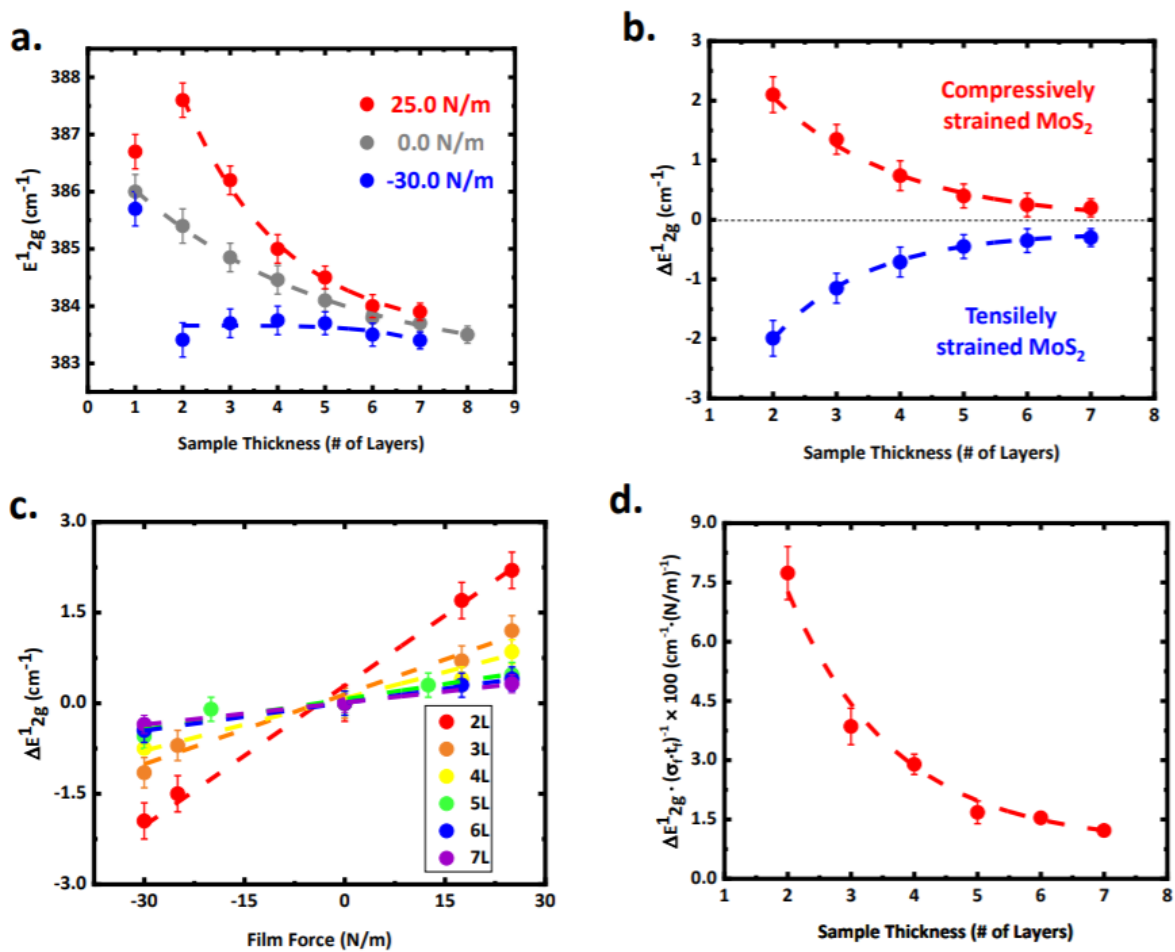

**Figure 2: (a)** Extracted $E^1_{2g}$ peak position trends varying with MoS$_2$ sample thickness. Red denotes tensile MgF$_2$ multilayer encapsulation, grey is no stressor, then blue is a compressive MgO multilayer encapsulations. Dashed lines exhibit fitted exponentials to each curve for clarity. **(b)** Calculated $E^1_{2g}$ peak shifts from (a). **(c)** Displayed is the $E^1_{2g}$ peak shifts for each thickness sample with varying thin film force. The dashed lines are the fitted linear function for each layer. **(d)** The slope for each layer (determined from (c)) is then plotted and fit to an exponential decay function (dashed line). Thin film stressors presented here are all e-beam evaporated multilayers.

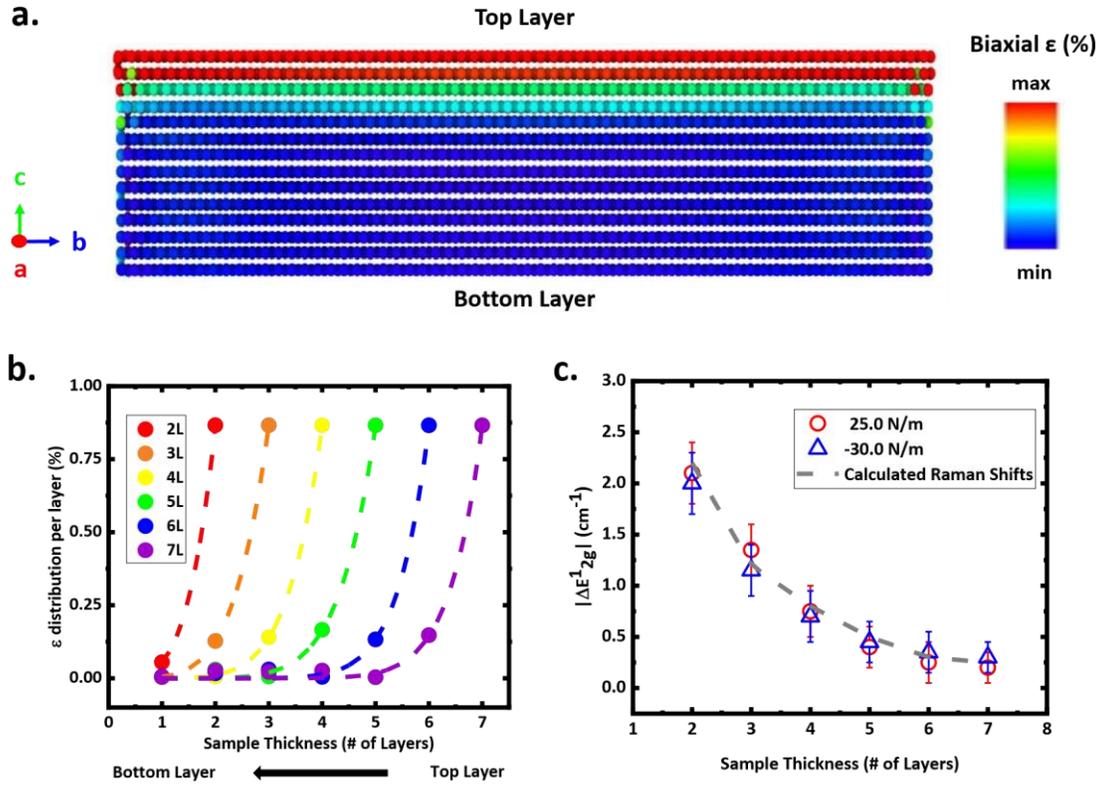

**Figure 3: (a)** Presents the strain distribution throughout a 7L sample of MoS$_2$ from MS simulations. Bottom layer is fixed, mimicking adhesion to the substrate. **(b)** Layer-by-layer strain distribution for various thickness samples determined from MS. **(c)** Presents calculated E$^1_{2g}$ peak shifts (grey line) compared to actual measured E$^1_{2g}$ peak shifts for both tensile and compressive thin film stressors. Note we take the magnitude of ΔE$^1_{2g}$ to directly compare the changes in the tensilely and compressively stressed MoS$_2$. Experimental Raman data is that presented in Fig. 2b.

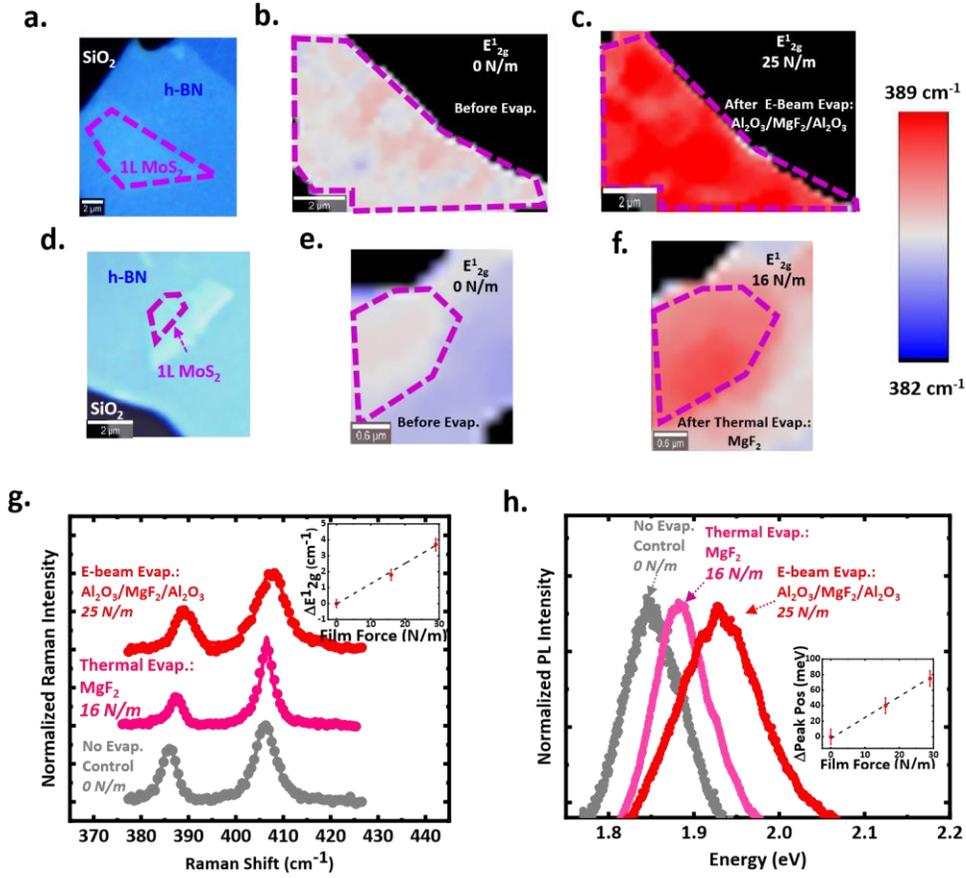

**Figure 4: (a)** Optical micrograph of the 1L-MoS$_2$/h-BN/SiO$_2$/Si sample presented in (b) and (c). **(b)** E$^1_{2g}$ peak position map of the flake from (a) with no encapsulation. **(c)** E$^1_{2g}$ peak position map of the flake from (a) after an e-beam evaporated tensile MgF$_2$ multilayer encapsulation. **(d)** Optical micrograph of the 1L-MoS$_2$/h-BN/SiO$_2$/Si sample presented in (e) and (f). **(e)** E$^1_{2g}$ peak position map of the flake from (d) with no encapsulation. **(f)** E$^1_{2g}$ peak position map of the flake from (d) after a tensile thermally evaporated MgF$_2$ alone encapsulation. **(g)** Averaged Raman spectra from the samples is presented. Grey spectrum shows an averaged spectrum with no encapsulation, pink shows that of the thermally evaporated MgF$_2$ encapsulated sample, and finally the red shows that of the e-beam evaporated MgF$_2$ multilayer encapsulated sample. Inset shows change in E$^1_{2g}$ peak position versus thin film force. **(h)** Photoluminescence (PL) signatures from the MoS$_2$ samples is presented. Grey spectrum shows a typical MoS$_2$ PL with no encapsulation, pink shows that of the thermally evaporated MgF$_2$ encapsulated sample, and finally the red shows that of the e-beam evaporated MgF$_2$ multilayer encapsulated sample. Inset shows change in primary PL peak position (A peak) versus thin film force. All Raman maps presented here follow the color bar on the right. Monolayer regions are outlined in a dashed purple lines.

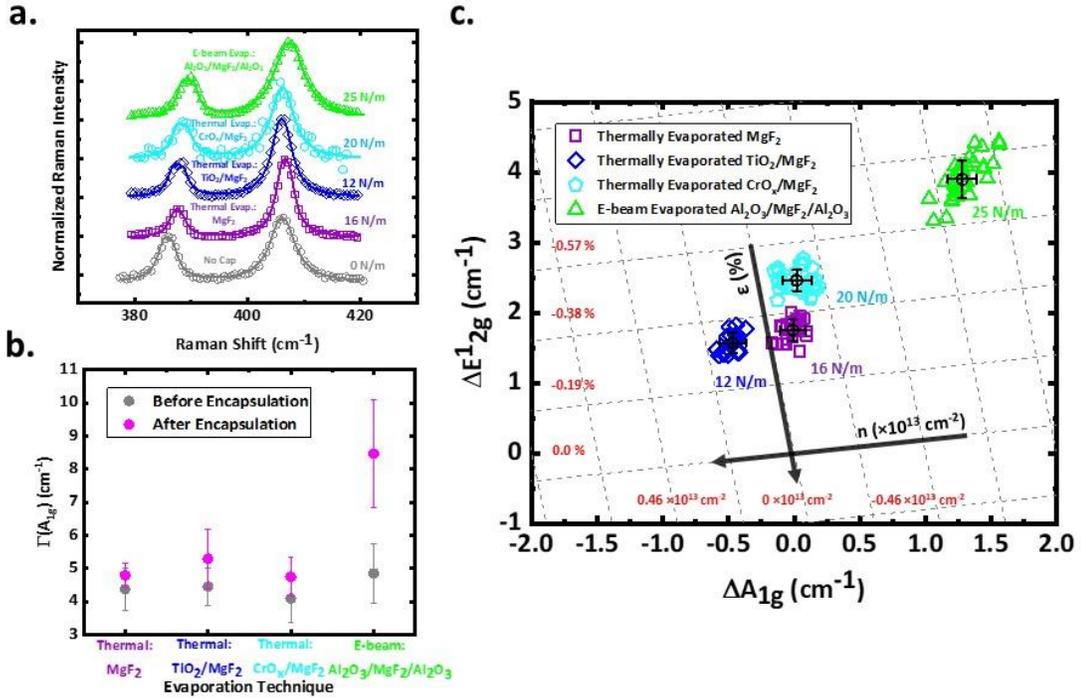

**Figure 5: (a)** Raman spectra of 1L-MoS$_2$/h-BN/SiO$_2$/Si samples with different tensilely stressed MgF$_2$ encapsulations. **(b)** A$_{1g}$ full-width-half maximum ($\Gamma$(A$_{1g}$)) comparison of before and after stressor encapsulations. Error bars correspond to standard error from the fitting procedure. **(c)** Change in A$_{1g}$ versus change in E$^1_{2g}$ peak positions, calculated from subtracting after encapsulation peak positions from before encapsulation. Grid lines present vector space for strain and doping. Major tick marks are accompanied by the quantified strain and electron concentration. Each cluster of data sets have the mean peak shifts values marked with an open circle symbol, provided with error bars. Vector space is constructed using previous detailed spatial Raman characterization work.[23]

# Supplementary Information:

# Strain Engineering 2D MoS$_2$ with Thin Film Stress Capping Layers

Tara Peña,*,† Shoieb A. Chowdhury,‡ Ahmad Azizimanesh,† Arfan Sewaket,†

Hesam Askari,‡ and Stephen M. Wu*,†,¶

†*Department of Electrical & Computer Engineering, University of Rochester, Rochester, NY, USA.*

‡*Department of Mechanical Engineering, University of Rochester, Rochester, NY, USA.*

¶*Department of Physics & Astronomy, University of Rochester, Rochester, NY, USA.*

E-mail: tpena@ur.rochester.edu and stephen.wu@rochester.edu

# Layer Identification

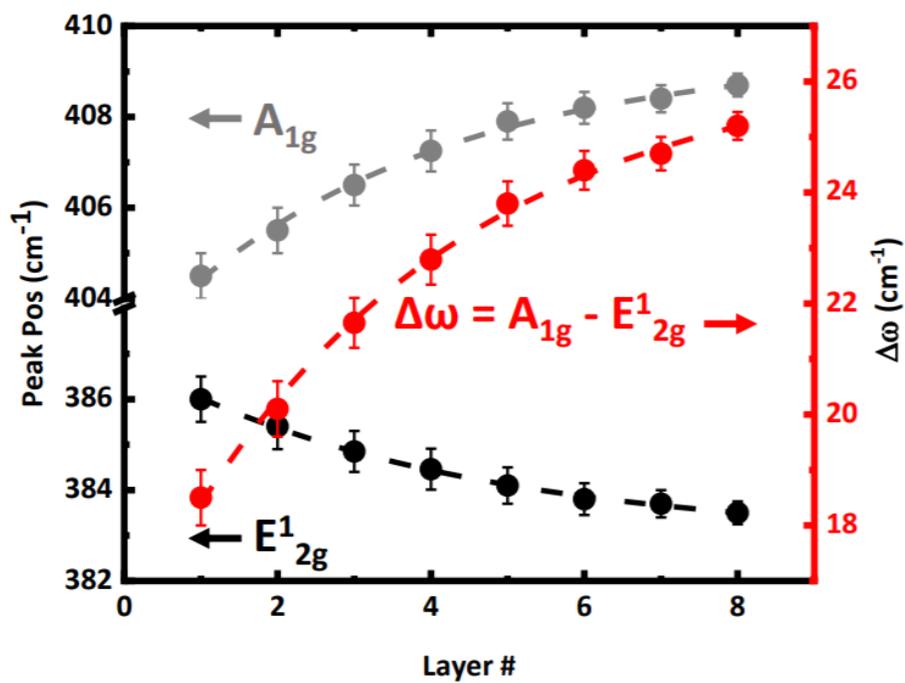

**Figure S1:** The in-plane (black) and out-of-plane (grey) Raman peak positions with respect to the number of layers for control $MoS_2$ on a single-crystal MgO substrate. Red curve denotes the difference between the two peaks.

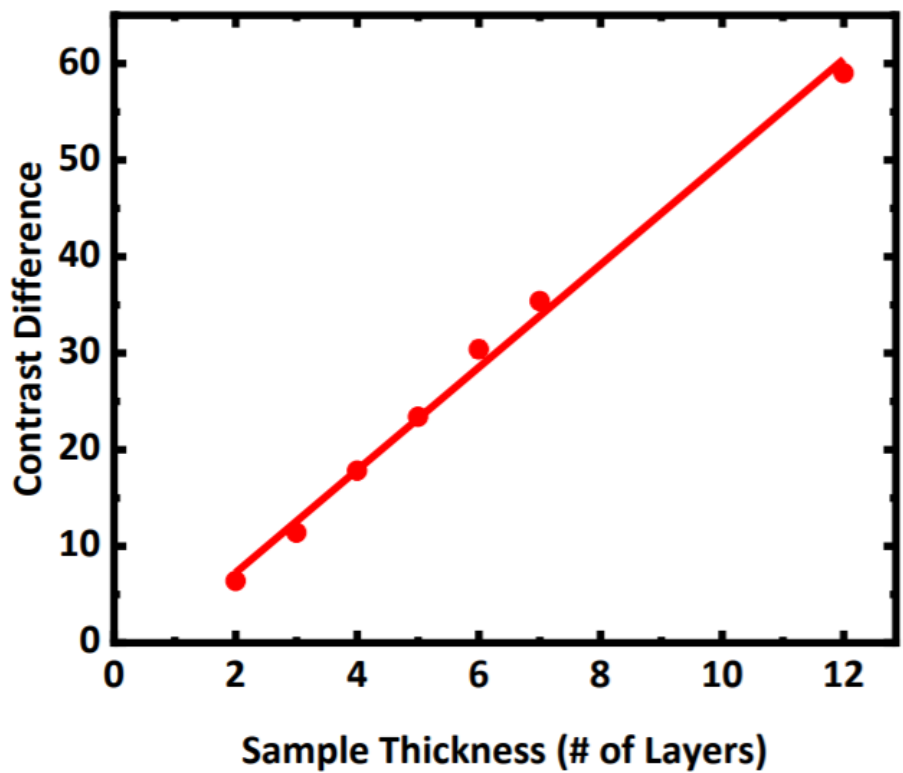

**Figure S2:** Optical contrast difference between MoS$_2$ layers and single-crystal MgO substrates. Contrast values are determined from optical micrographs with same exposure.

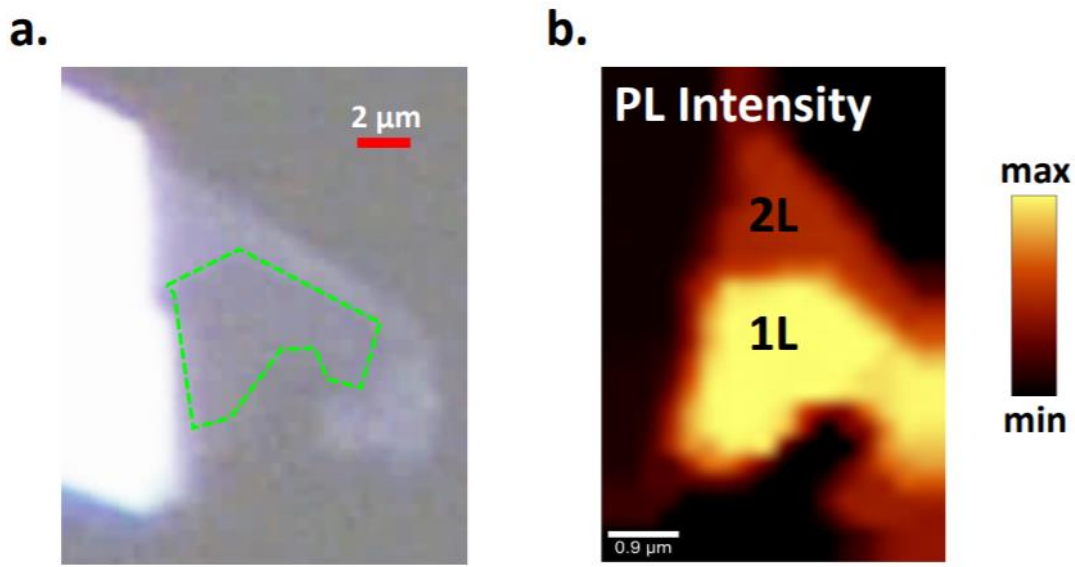

**Figure S3:** Monolayer identification. **(a)** Control monolayer (1L) and bilayer (2L) MoS$_2$ optical micrograph, green dashed line outlines the monolayer region. **(b)** Photoluminescence (PL) intensity map of the same sample, confirmation of monolayer is made by the drastic increase of intensity and an optical bandgap of 1.85 eV.

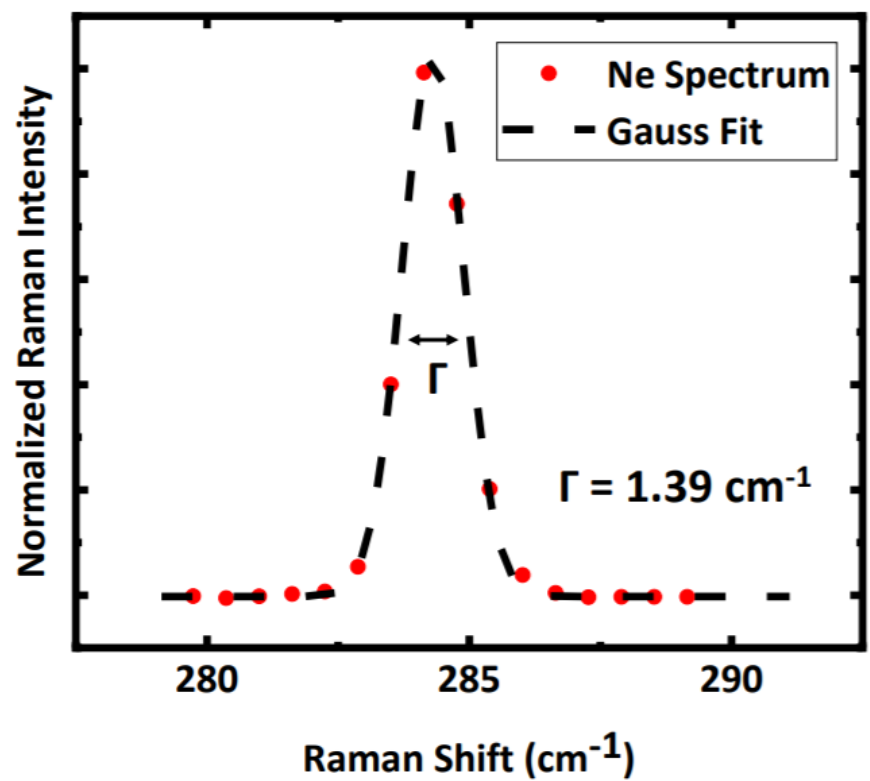

**Figure S4:** Neon lamp spectral line example, demonstrating a Gaussian Γ of 1.39 cm⁻¹. This parameter is essential to decouple instrumental contributions from the intrinsic $MoS_2$ response.[1] All fittings in this work were done using a Voigt profile, fixing the Gaussian Γ of that found experimentally.

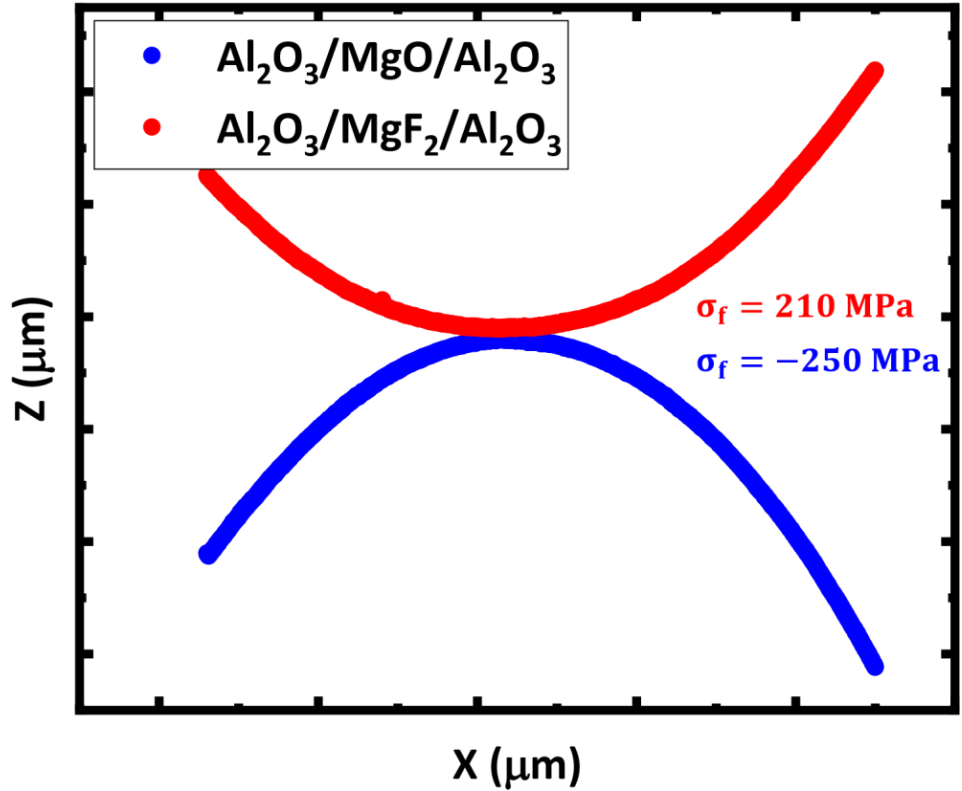

**Figure S5:** Surface profilometer data for both tensile (MgF₂) and compressive (MgO) e-beam evaporated thin film stressor multilayers, stress was calculated using the Stoney equation. All e-beam evaporated films presented in this work are multilayers following: (10 nm) $Al_2O_3$ / X / (10 nm) $Al_2O_3$, where X = $MgF_2$, MgO, or $SiO_2$.

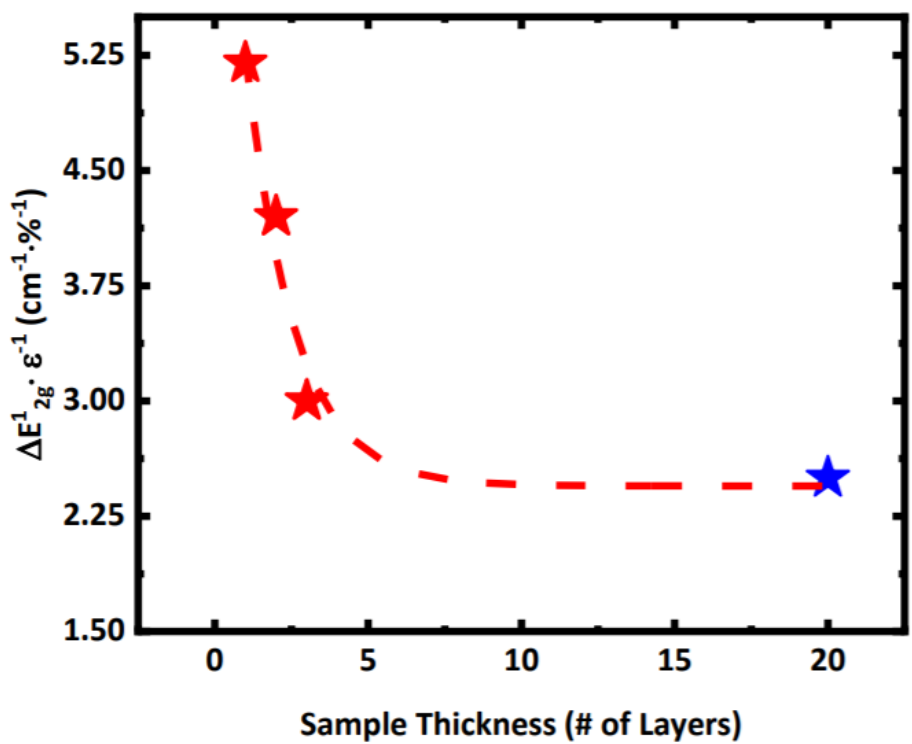

**Figure S6:** For $MoS_2$ under biaxial in-plane strain, there are linear translation factors that relates $E^1_{2g}$ peak shifts to strain magnitude. This translation value has been experimentally determined to vary with $MoS_2$ thickness. We estimate these factors from previous work implementing biaxial strain onto $MoS_2$. Red stars denote translation factors determined from previous biaxial strain work.[2] Blue star denotes translation factor value for bulk $MoS_2$.[3] Dashed line represents exponential decay fit, extrapolating these values for layers 4-7.

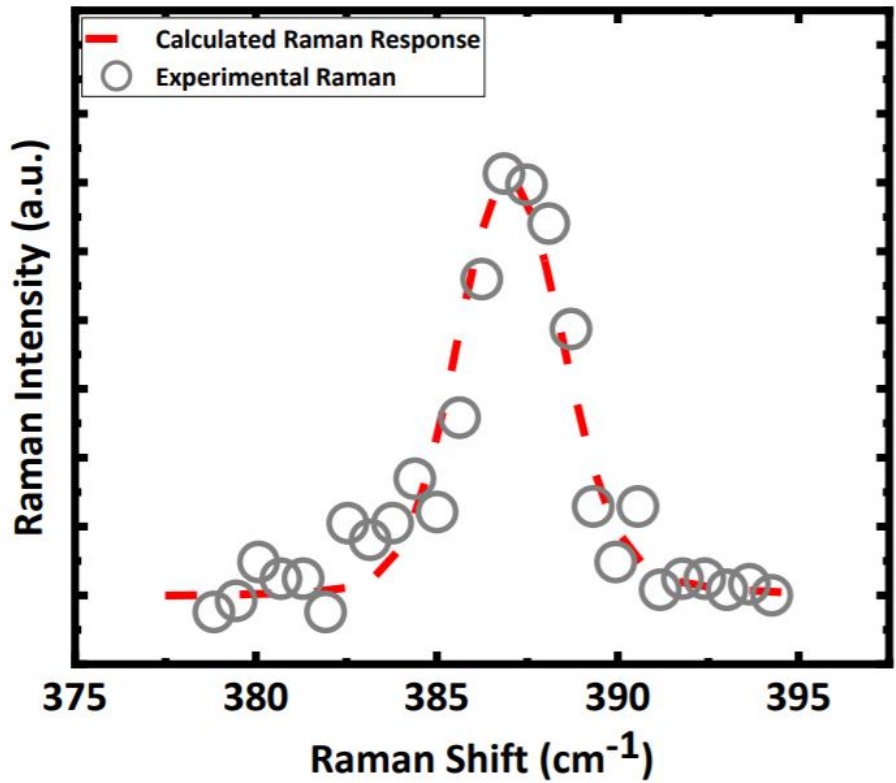

**Figure S7:** Example of a calculated $E^1_{2g}$ Raman response for a 2L compressively strained $MoS_2$ sample. Two Lorentzian functions with peak positions 389.1 $cm^{-1}$ (top layer, 0.85% strained) and 385.85 $cm^{-1}$ (bottom layer, 0.08% strained) were superimposed. From this superposition, a peak position of ~387.5 $cm^{-1}$ was determined. The calculated Raman response presented is the superimposed Lorentzians convolved with a Gaussian response ($\Gamma$ = 1.39 $cm^{-1}$), to replicate what we observed experimentally. Open circles are from experimental Raman data of a compressively strained 2L $MoS_2$ sample.

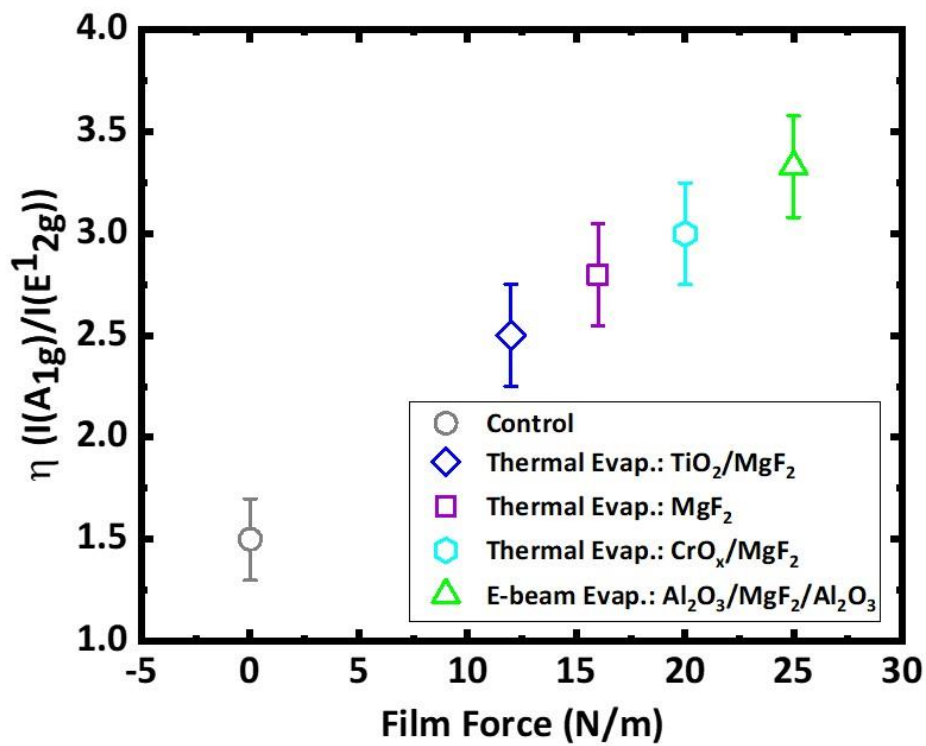

**Figure S8:** We analyze the intensity ratio between the $A_{1g}$ and $E^1_{2g}$ peaks. The spectra used are from 1L-MoS$_2$/h-BN/SiO$_2$/Si samples with the same thin film stressors presented within the main text. Increase of this ratio, $\eta$, matches theoretical work with compressive biaxial strain onto 1L TMDCs.[4]

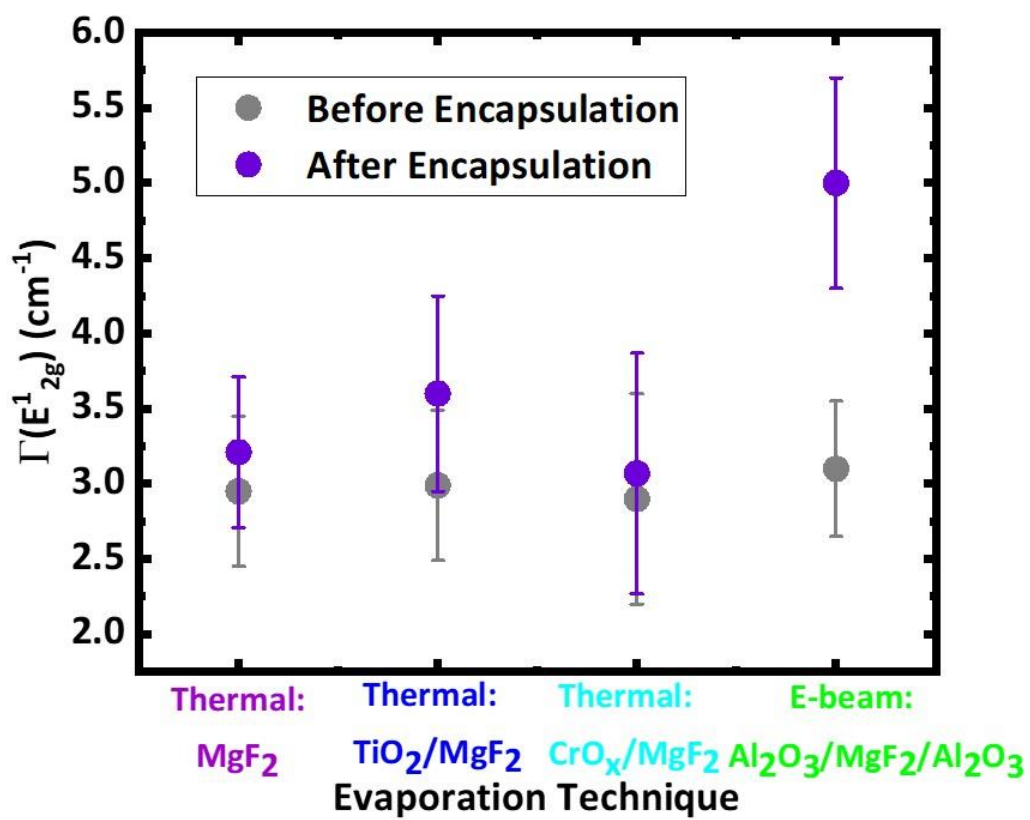

**Figure S9:** $E^1_{2g}$ full-width-half maximum ($\Gamma(E^1_{2g})$) comparison of before and after stressor encapsulations. Error bars correspond to standard error from the fitting procedure.

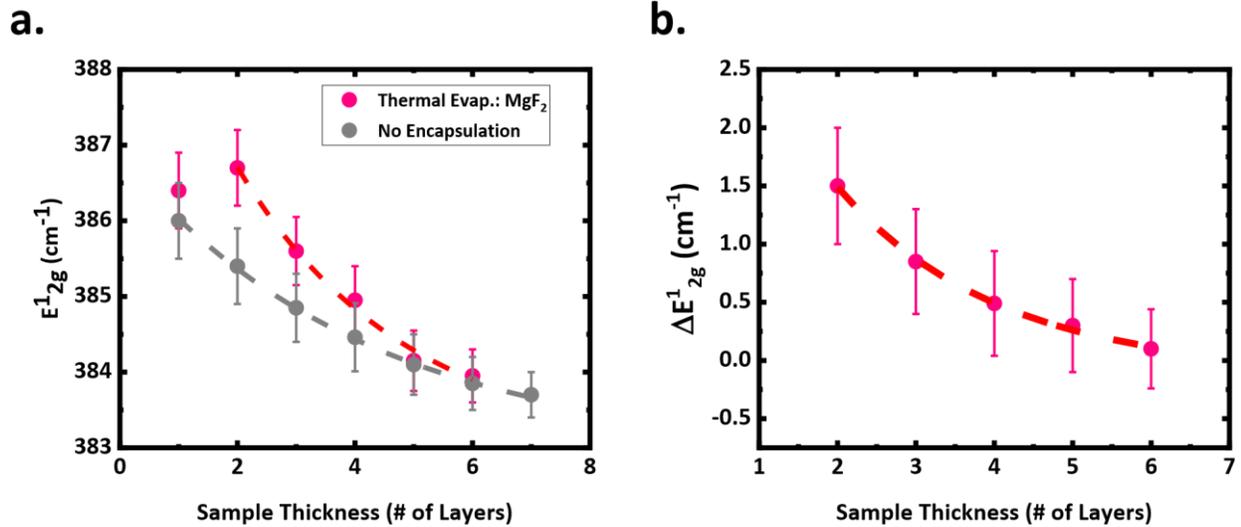

**Figure S10:** **(a)** Extracted $E^1_{2g}$ peak position trends varying with $MoS_2$ sample thickness. Pink denotes thermally evaporated $MgF_2$ alone encapsulation (16 N/m film presented in the main text) and grey denotes no encapsulation (control). **(b)** Calculated $E^1_{2g}$ peak shifts from thermally evaporated $MgF_2$ alone versus control samples presented in (a). All dashed lines are fitted exponentials to each curve for clarity.

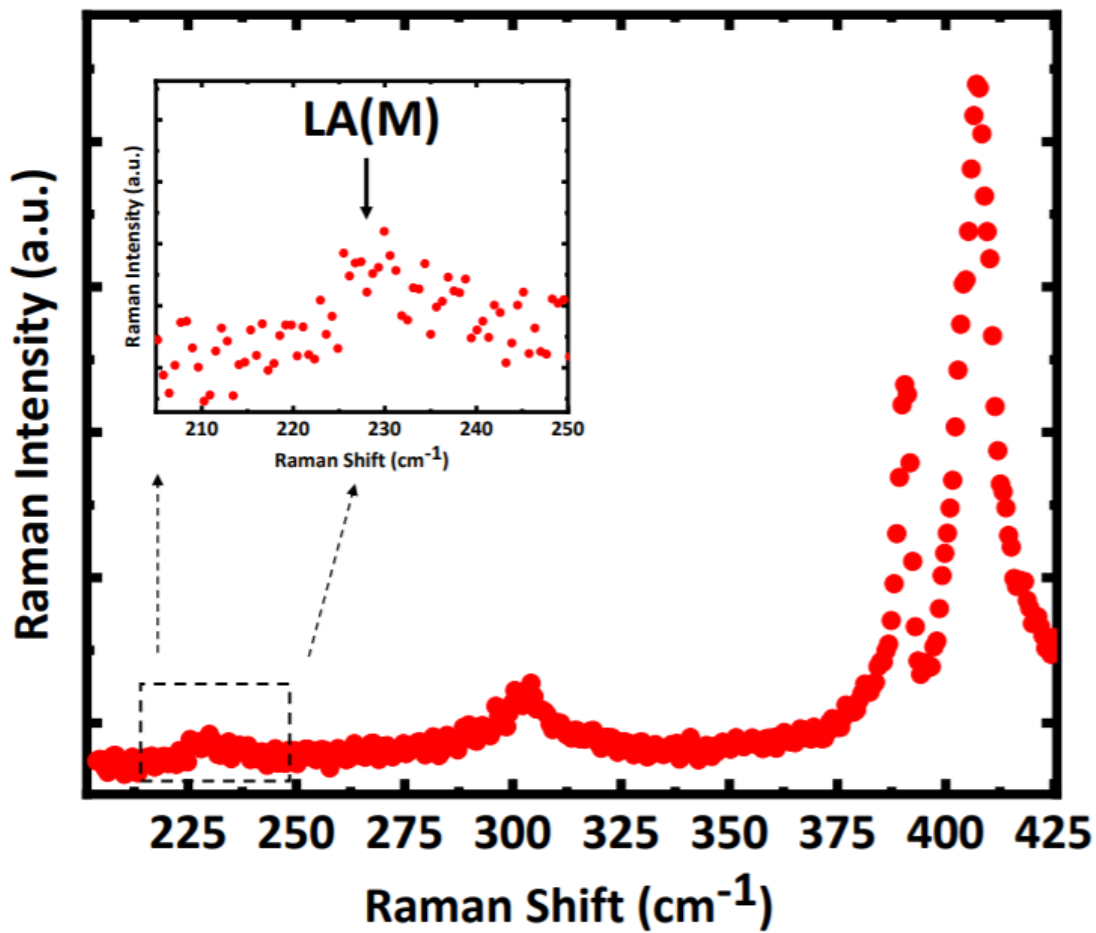

**Figure S11:** Single spectrum of the e-beam evaporated $MgF_2$ multilayer encapsulated sample of 1L-$MoS_2$/h-BN/$SiO_2$/Si presented in main text. This is the only sample we found the LA(M) peak after encapsulation, this peak location is ~227 $cm^{-1}$. The LA(M) peak has been associated with defects within $MoS_2$.[5] One may extract the interdefect length ($L_D$) of a given sample by analysis of the intensity ratios I(LA(M))/I($A_{1g}$) and I(LA(M))/I($E^1_{2g}$). Using the analysis from Mignuzzi *et. al.*, we extract a $L_D$ of ~3.9 nm. The other three 1L-$MoS_2$/h-BN/$SiO_2$/Si samples with thermally evaporated encapsulations did not present LA(M) peaks.

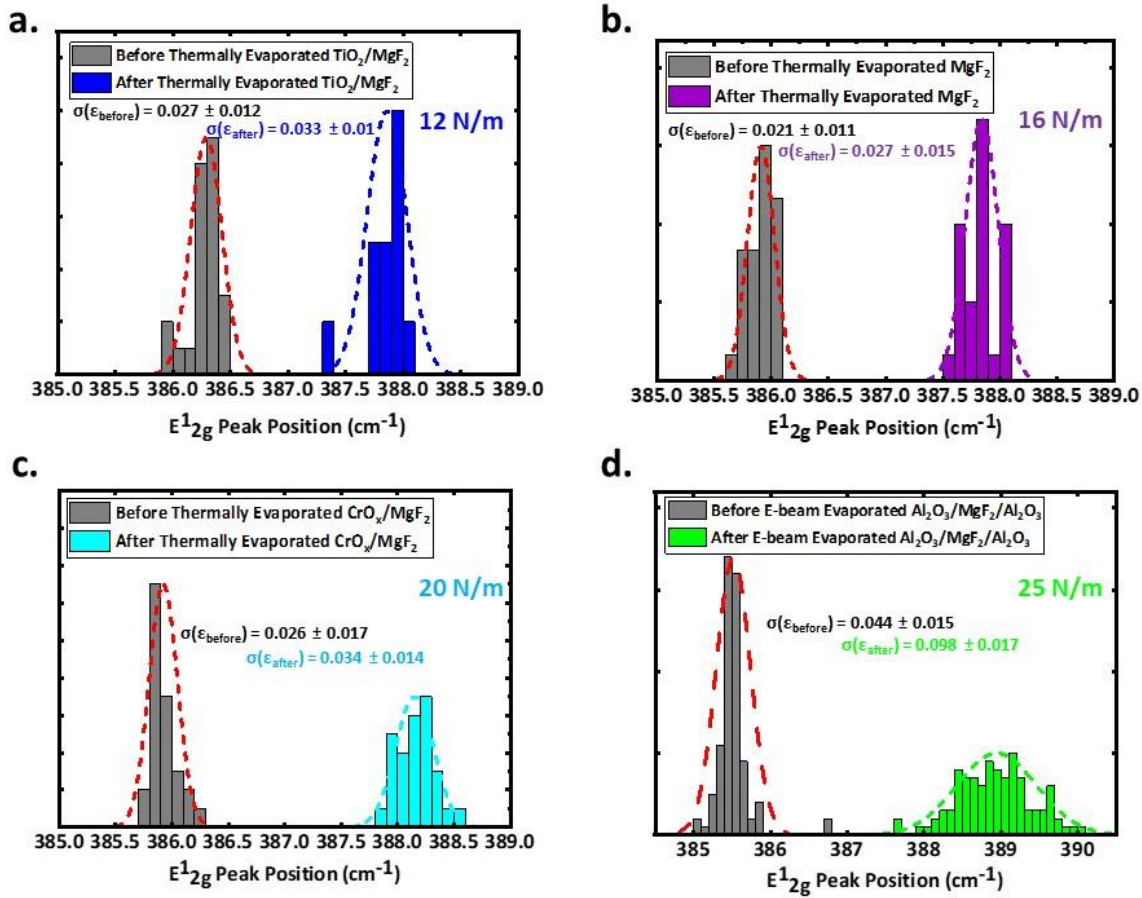

**Figure S12:** Histograms of the $E^1_{2g}$ peak positions before and after thin film stressor encapsulations of the 1L-$MoS_2$/h-BN/$SiO_2$/Si samples presented in the main text. **(a)** Thermally evaporated $TiO_2$/$MgF_2$ sample. **(b)** Thermally evaporated $MgF_2$ alone sample. **(c)** Thermally evaporated $CrO_x$/$MgF_2$ sample. **(d)** E-beam evaporated $Al_2O_3$/$MgF_2$/$Al_2O_3$ sample. We find the sample with the most variation after encapsulation is that in the e-beam evaporated $Al_2O_3$/$MgF_2$/$Al_2O_3$ sample, most likely from the stronger presence of defects/doping introduced into this sample.

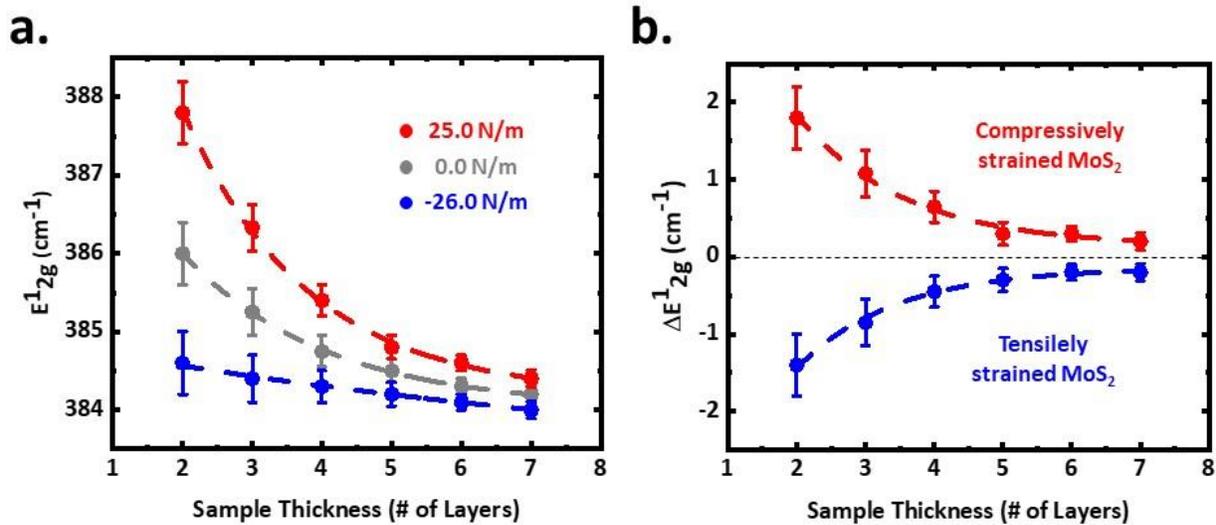

**Figure S13:** Thin film stressor technique on $MoS_2/SiO_2/Si$ samples. We utilize atomically flat single-crystal MgO substrates in Figures 1-3, however one may replace the MgO for any 3D-bonded substrate with a reasonable surface roughness. The $SiO_2/Si$ substrate employed here has a $R_a = 0.3$ nm. **(a)** $E^1_{2g}$ peak position versus thickness for varying thin film forces. **(b)** The calculated $\Delta E^1_{2g}$ determined from (a) versus thickness of the two thin film forces. Here we confirm the same exponential decay trends that we observe in Fig. 2.